# A Bioinformatics Study for Recognition of Hub Genes and Pathways in Pancreatic Ductal Adenocarcinoma

**Atefeh Akbarnia Dafrazi[1], Tahmineh Mehrabi[2], Fatemeh Malekinejad[3]**

[1]Department of Chemistry, Faculty of science, University of guilan, Rasht, Iran

[2]Department Name, Name of Institution, City, Country (second author's affiliation)

[3]Department of Animal Biology, School of Biology, University College of Science, University of Tehran, Tehran, Iran

**Corresponding Author's information:** Atefeh Akbarnia Dafrazi, e-mail: atefe.aka@gmail.com

**Source(s) of financial support** : Nothing

**Conflict of interest** : None

## Abstract

**Background:** The aim of this study is to use bioinformatics to discover the biomarkers associated with patients with PDCA (one of the deadliest malignancies worldwide).

**Material and Methods:** GSE28735, GSE15471, and GSE62452 are gene microarray datasets drived from the GEO database, included 153 Pancreatic ductal adenocarcinoma samples and 145 normal samples. By analyzing both Gene Ontology (GO) and the Kyoto Encyclopedia of Genes (KEGG), for screening DEGs has provided information about their biological function. Protein-protein interactions (PPI) of DEGs were analyzed using the Search Tool for the Retrieval of Interacting Genes database (STRING) and visualized by Cytoscape. UALCAN was also used to perform prognostic analyses.

**Results:** In Pancreatic ductal adenocarcinoma, we discovered 2264 upregulated DEGs (uDEGs) and 723 downregulated DEGs (dDEGs). The Gene Ontology (GO) indicated that extracellular matrix organization, extracellular structure organization, collagen catabolic process and also the uDEGs are enriched in focal adhesion, PI3K-Akt signaling pathway, ECM-receptor interaction by KEGG analysis. The top ten hub genes were identified from the PPI network: COL1A1, COL3A1, COL1A2, FN1, COL5A2, ITGA3, FBN1, MET, COL6A3, and BGN. These 10 hub genes are highly upregulated in pancreatic ductal adenocarcinoma, according to GEPIA research. The UALCAN prognostic analysis of the 10 hub genes revealed that two of the elevated genes, ITGA3 and MET, significantly shortened PDAC patient survival time. Pancreatic ductal adenocarcinoma is linked to Focal adhesion, the PI3K-Akt signaling pathway, and ECM-receptor interaction, according to Module analysis from the PPI network.

**Conclusion:** This investigation identified hub genes and important signal pathways, which adds to our knowledge of molecular mechanisms and could be exploited as diagnostic and therapeutic biomarkers for Pancreatic ductal adenocarcinoma.

**MeSH Keywords:** Diagnosis, Hub genes, Interactions, Pathways

**Background:**

Pancreatic ductal adenocarcinoma (PDAC) is one of the most deadly and aggressive malignancies in the world. It forms Almost 90% of all pancreatic cancers (PCs) [1, 2]. Because of the late diagnosis, persons with this cancer have a short life expectancy. The disease's signs can be hidden until late stage, where the 5-year survival rate is below 3%[3].

Many scientists are investigating new biomarkers for identification and targeted cancer therapy via gene differentially gene expression on a large scale via genetic chip, which is known as high-throughput sequencing [4]. Therefore, it is necessary to study these new diagnostic markers in pancreatic adenocarcinoma in order to help early diagnosis of this disease and prevent its growth and progression. since the advent of bioinformatical methods and the rapid-paced development of technology massive data from genetic alterations during the onset and course of the disease is highly used [5]. Microarray technology is well-designed to recognize the variety of genes, which gradually becomes the infrastructure of diagnosis and treatment [5].

GEO is a publicly accessible data repository for functional genomics that allows for data mining of cancer gene expression profiles [6].

This study investigated the dissimilarly expressed genes (DEGs) in 3 PDAC datasets using the Geo database. Then to obtain differentially expressed genes (DEGs), we employed GEO2R, a bioinformatics method based on the R programming language. In order to enrich analysis of biological function and pathway, we used Gene Ontology (GO) and Kyoto Encyclopedia of Genes and Genomes (KEGG) for 1006 overlap genes which gained among the three datasets. In addition, Search Tool for the Retrieval of Interacting Genes (STRING) and Cytoscape were utilized to extract a comprehensive view of protein-protein interactions (PPI). Following this, we use UALCAN for predicting prognosis and GEPIA for verification of mRNA levels for the purpose of conducting molecular studies on more understanding of mechanism of PDAC. This will help diagnose PDAC, targeted drug research, and evaluate its prognosis.

**Material and Methods:**

**The analysis of microarry data**

There is a free database available called Gene Expression Omnibus (GEO) that includes array-based and sequence-based data, and can be found at http://www.ncbi.nlm.nih.gov/geo). Data pertaining to gene expression profiles GSE28735 [7], GSE15471 [8], and GSE62452 [9] of PDAC are all derived from the GEO database. For this experiment, three datasets were chosen: Samples from human PDAC tissue; cases with a control group; and Specifically, sample number ≥ 18, and just for PDAC related to pathological characteristics. GSE28735 was based on the GPL6244 platform ([HuGene-1_0-st] Affymetrix Human Gene 1.0 ST Array [transcript (gene) version]). GSE15471 was based on the GPL570 platform ([HG-U133_Plus_2] Affymetrix Human Genome U133 Plus 2.0 Array). GSE62452 was based on the GPL6244 platform ([HuGene-1_0-st]

Affymetrix Human Gene 1.0 ST Array [transcript (gene) version]). Dataset GSE28735 included 90 samples, 45 Pancreatic ductal adenocarcinoma samples and 45 normal samples. Dataset GSE15471 included 78 samples, 39 Pancreatic ductal adenocarcinoma samples and 39 normal samples. Dataset GSE62452 included 130 samples, 69 Pancreatic ductal adenocarcinoma samples and 61 normal samples.

**Screening of DEGs**

A Web tool called GEO2R (https://www.ncbi.nlm.nih.gov/geo/geo_HYPERLINK applied to compare distinct GEO serieses and to identify DEGs. In order to analyze the data, GEO2R was divided into a Pancreatic ductal adenocarcinoma group and a normal group [10]. Each dataset was statistically analyzed using adj. $p<0.05$ and $|log2FC|> 1$ and The intersection of the three datasets was determined by using the online tool Draw Venn diagram (bioinformatics. psb. ugent. be/webtools/Venn).

**Gene ontology (GO) and KEGG signal pathway enrichment analysis of DEGs**

The Gene Ontology (GO) is quite possibly the most broadly utilized theoretical assets for allocating functional ascribes to genes and genic productions [11]. The GO (http://www.geneontology.org) information base can give genomic information, including organic cycles (BP), cell part (CC), and sub-atomic capacity (MF) [12].

A networked website called the Kyoto Encyclopedia of Genes and Genomes (KEGG) including biological systems can be found at http://www.genome.ad.jp/kegg/) database [13] and can be provided information about genic functional and its analysis and interpret.

An online database as The Database for Annotation, Visualization and Integrated Discovery (DAVID, http://david.abcc.ncifcrf.gov/) applies for gene probing to distinguish the biological performance [14]

In this study, GO and Kegg analyses with P<0.05 to demonstrate a statistical significance were utilized for investigating the performances of DEGs.

**Integration of PPI network and module analysis**

The Search Tool for the Retrieval of Interacting Genes (STRING,*http://string.embl.de/*) [15] is a designed biological database to provide predicting PPI networks.

We assessed the interactive relationships by importing DEGs into STRING and assuming a confidence score of >0.7 [15]. Following that, we constructed predictive models using a system called Cytoscape [16], which allows visualizing biological molecular interactions.

A node's degree is determined by the number of edges it has connected to the special node. Therefore nodes with a high degree were known as hub genes (contributing to essential biological functions).The benchmarks were determined as: degree cutoff=2, node score cutoff=0.2, k-core=4, and maximum depth=100.

**Expression levels and prognostic analysis of hub genes**

The Gene Expression Profling Interactive Analysis (GEPIA) [17] provides users with access to a platform for analyzing differences between mRNA expression levels of a specifc gene in special cancers between cancerous tissues and paired normal tissues. PDAC and paired normal tissues were analyzed by GEPIA for the expression levels of hub genes. The prognosis of hub genes was assessed using UALCAN (http://ualcan.path.uab.edu) [18]. The values of RNA expression were according to the tumor types of the patients, however, $p<0.05$ was assumed significant for each gene.

**Results**

**Screening of DEGs**

In altogether, 3 datasets consisted of a total of 298 samples, 153 pancreatic ductal adenocarcinoma samples, and 145 matched normal samples were analyzed. According to the GEO2R analysis, by Applying the adj. p<0.05 and |log2FC|>1 criteria, a total of 2987 DEGs were selected, which consisted of 2264 upregulated DEGs (uDEGs) and 723 downregulated DEGs (dDEGs) were screened in PDAC tissues compared with normal tissues (Figure 1). 3 datasets, including 147 uDEGs (Figure 2, Table 1) and 55 dDEGs (Figure 2, Table 1) as a total of 202 genes were collected.

**GO term enrichment analysis**

The GO analysis outcomes demonstrated that the markedly enriched based on biological processes (BP) in extracellular matrix organization, extracellular structure organization and collagen catabolic process (Figure 3, Table 2), while the dDEGs are significantly involved in proteolysis, chemical homeostasis, and peptide catabolic process (Figure 3, Table 2). Also molecular function (MF) indicated that the uDEGs are enriched in extracellular matrix structural constituent, glycosaminoglycan binding, and integrin binding (Figure 3, Table 2); and the dDEGs are enriched in exopeptidase activity, peptidase activity, acting on L-amino acid peptides, and peptidase activity (Figure 3, Table 2). Beside of these, the upregulated DEGs for the Cellular component (CC) analysis are concentrated in extracellular region part, extracellular region, and extracellular matrix (Figure3, Table 2), and dDEGs are focused in extracellular region, extracellular region part, and extracellular exosome (Figure 3, Table 2).

**KEGG signal pathway analysis**

As can be seen in table 3, the most significantly enriched pathways of uDEGs and dDEGs reconnoitered by KEGG analysis. The uDEGs are enriched in focal adhesion, PI3K-Akt signaling pathway, ECM-receptor interaction, while the dDEGs are enriched in Pancreatic secretion, Complement and coagulation cascades, signaling pathway.

**PPI network construction, module analysis, and hub genes determination**

The gene expression of 202 DEGs was calculated using the STRING database which is the online tool to determine calculation and interaction between proteins. Then DEGs were arrived into Cytoscape software for visualization. PPI network demonstrates 202 nodes and 174 edges which are involved in this network (Figure 4). From the PPI network, The Top10 genes with the highest connectivity were filtered as hub genes. The results displayed that COL1A1 was the highest rank among all DEGs, with 24 degree, following that COL3A1, COL1A2, FN1, COL5A2, ITGA3, FBN1, MET, COL6A3, and BGN were found (Table 4).

The analysis of 202 nodes in Cytoscape software showed that the core module with higher score involves 17 nodes and 174 edges (Figure 4) which reveals the critical role of all 17 nodes. They are all upregulated genes in Pancreatic ductal adenocarcinoma. The results of the KEGG signal analysis demonstrated the 11 genes of the 17 genes in the module mainly participated in three pathways: Focal adhesion, PI3K-Akt signaling pathway and ECM-receptor interaction. Three results showed that 11 genes including (COL1A1, COL3A1, COL1A2, LAMB3, ITGA3, COL11A1, LAMA3, COL5A2, FN1, COL6A3 and LAMC2) enriched in three pathways.

**Expression levels and prognostic analysis of hub genes**

The analysis of GEPIA indicated that these 10 hub genes are significantly upregulated in pancreatic ductal adenocarcinoma (Figure 5). We applied UALCAN to show the prognostic value of 10 hub genes. The analysis of prognostic showed 2 of the upregulated genes include ITGA3 and MET are significantly reduced the survival time of PDAC (Figure 6).

**Discussion**

Pancreatic ductal adenocarcinoma (PDAC) is the most common malignant tumor with a low survival rate. Recently, an early diagnosis to predict the potential biomarkers is applied to predict the potential biomarkers. Therefore, in this study, we explored candidate biomarkers genes and pathways of PDAC [19].

In the present study, 2264 upregulated DEGs (uDEGs) and 723 downregulated DEGs (dDEGs) were identified from GSE28735, GSE15471, and GSE62452 datasets by downloading from the GEO database, and then 147 uDEGs and 55 dDEGs were effectively expressed. we used a series of bioinformatics tools toprognosticand definition of the role of significant DEGs in pancreatic ductal adenocarcinoma.

The Gene Ontology (GO) shown that extracellular matrix organization, extracellular structure organization, collagen catabolic process while the dDEGs are in proteolysis, chemical homeostasis, and peptide catabolic process. evidence has shown that the ECM may be tumor-promoting [20]. and The deposition of abundant ECM proteins is common among solid tumors like PDAC and is known as a desmoplastic reaction [21].

For MF, the uDEGs are markedly enriched in extracellular matrix structural constituent, glycosaminoglycan binding, and integrin binding, while the dDEGs were enriched in exopeptidase activity, peptidase activity, acting on L-amino acid peptides, and peptidase activity.

GO CC analysis revealed that extracellular region part, extracellular region, and extracellular matrix, while dDEGs were concentrated in extracellular region, extracellular region part, and extracellular exosome.

The top 10 hub genes were screened out by Cytoscape software and constructed a PPI network of protein-protein interaction between uDEGs. The top 10 hub genes from high to low degree was COL1A1, COL3A1, COL1A2, FN1, COL5A2, ITGA3, FBN1, MET, COL6A3, and BGN,
which belong to the collagen (COL) family, are the top 10 hub genes, which suggests that the collagen gene is likely to be a potential target for pancreatic ductal adenocarcinoma. The entire family contains 19 types of collagen and more than ten types of collagen-like proteins and is encoded by more than 30 different genes [22]. Due to differences in its molecular structure, collagen can be divided into two categories: fibrogenic collagen and nonfibrogenic collagen [22] Types I, II, III, V, and IX collagen are fibrogenic [22]. The extracellular matrix (ECM) is composed primarily of collagen, which is a macromolecular substance that supports the cell structure and regulates the physiological activities of the cell [23]. ECM degradation is one of the most important steps that leads to cancer cell invasion and metastasis [24]. Stably expressed collagen is necessary to maintain the normal functions of cells and tissues. However, the abnormal overexpression of collagen is associated with a variety of pathological processes, especially in malignant tumors.

COL1A1 is a member of the type I collagen family and encodes the pro-a 1 chains of type I collagen [25]. Yang et al. [26] found that the miRNA sponge hsa-circRNA-0007334 inhibits hsa-mir-577 expression, leading to the overexpression of themir-577 target gene COL1A1and subsequently causing PDACcell migration. They also demonstrated that high COL1A1 shortens the survival time of PDAC patients [26].

COL3A1 is a type III collagenase whose lack causes perforation, tearing, fracture and even fragmentation of connective tissue-related structures in the body [27]. Hall et al. [28]revealed that the overexpression of COL3A1 in patients with pancreatic cancer can be significantly downregulated after gemcitabine combined with EC359 treatment.

ECM is a protein compound that plays an indispensable role in cell migration and cancer development [29].

At present work, it is reported that COL1A2 involves in the regulation of osteogenesis imperfecta (OI)[30], Ehlers-Danlos syndrome [31], osteoarthritis[32], and so on. Furthermore, some researchers reported that COL1A2 might directly play a role in pancreatic cancer proliferation, migration, invasion, and in vivo xenograft progression, through its interaction with microRNA (miRNA) [33, 34]

The COL5A2 gene encodes a 46 kDa nuclear localization transcriptional inhibitor protein that has been reported to affect cancer progression [35]. Fischer et al. [36]illustrated that in normal colon tissue, COL5A2 is not expressed, but in colon cancer tissues, COL5A2 is expressed. Chen et al. [37]showed that in osteosarcoma, COL5A2 expression can be repressed by the tumor suppressor gene NKX2-2. Thus, these two genes are upregulated in other cancers and have a great impact on the progression of those cancers, indicating that they may also affect the progression of PDAC in the same manner.

COL6A3, as a type VI collagen a chain can interact with various components in the ECM [38]. Its abnormal expression in various cancers suggests that COL6A3 affects cancer formation. Svoronos et al. [39] showed that in the PDA stroma, COL6A3 is the major overexpressed gene, and overexpressed COL6A3 leads to a poor prognosis in PDAC. Arafat et al. [40] also showed that a high level of COL6A3 expression is found in the tissues of PDAC patients, especially those in later disease stages and that patients in earlier stages present relatively low COL6A3 levels.

Fibronectin 1 (FN1), belonging to the ECM glycoprotein family, functions in the ECM process and contributes to cellular adhesion, polarity, migration, and tissue remodeling [41, 42]. Besides, FN1 executes its functions in infection resistances, and microvascular integrity maintenance [43] High FN1 expression is associated with various human carcinomas, including pancreatic [44].

Studies have discovered that FN stimulates invasion and adhesion and markedly inhibits cell death of pancreatic cancer cells [45].

Our study also found the FN1 is enriched in ECM-receptor interaction at the KEGG signal pathway analysis.

The ITGA3 gene encodes integrin alpha-3, a component of the integrin family of proteins. ITGA3 is located on the cell membrane and functions as a cell surface adhesion molecule. A previous study reported that ITGA3 interacts with extracellular matrix proteins, including members of the laminin family, and that its expression correlates with cancer metastasis [46].

In addition, many omics studies have identified ITGA3 as a key molecular marker in several cancers, [47, 48], although its value in clinical settings is uncertain.

Previous studies reported that ITGA3 has a role in pancreatic duct adenocarcinoma [49].

The previous study shown that high ITGA3 expression was an autonomous prognostic factor for poor overall survival and relapse-free survival.[50] Thus, ITGA3 expression appears to be a useful diagnostic and prognostic biomarker for pancreatic cancer. ITGA3 serves as a diagnostic and prognostic biomarker for pancreatic cancer. A previous study found that ITGA3 may facilitate cancer development by activating the PI3K-Akt signaling pathway, which increases proliferation, migration, and invasion in many cancers [50].

FBN1 encodes fibrillin, which is the primary component of microfibrils in the extracellular matrix. A previous study reported that FBN1 overexpression plays a key role in the development of germ cell tumors [51]. FBN1 is also a target gene for microRNA- (miR-) 133b. miR-133b inhibits the proliferation, migration, and invasion of gastric cancer cells by increasing FBN1 expression [52]. Moreover, hypermethylated FBN1 is found in tissue samples from colorectal cancer patients but not in healthy controls, suggesting that hypermethylated FBN1 may be a sensitive biomarker for this disease [53]The role of FBN1 in the development of pancreatic cancer and its correlation with the prognosis of patients has not been reported previously

The protooncogene c-met encodes a member of the family of receptor tyrosine kinases that is a 190-ku glycoprotein comprised of a transmembrane 145-ku β subunit and an extracellular 50-ku α subunit [54, 55]. The met receptor binds to and is activated by hepatocyte growth factor/scatter factor (HGF/SF) [56], leading to increased proliferation, altered motility, and enhanced invasion [57]. The wild-type c-met gene is amplified or overexpressed in many types of human cancer, including cancers of the breast, stomach, liver, endometrium, nasopharynx, and pancreas [58, 59]. MET overexpression in patients with pancreatic cancer, and is associated with a poorer overall survival [60].

Biglycan (BGN; also known as proteoglycan-1 and dermatan sulfate PG-1) is a single-copy gene localized on the long arm of human X chromosome Xq13-qter [61]. This gene contains at least two introns and it spans ~6 kb in length [62] BGN is a key member of the small leucine-rich proteoglycan family that resides at the cell surface or in the pericellular space of tissues [63]. BGN is typically expressed in the nerve, bone, cartilage, skin and muscles, modulating the morphology, growth, adhesion, bone mineralization, inflammation, migration and differentiation of epithelial cells [64]. The upregulation of BGN has been reported in multiple types of solid cancer, including ovarian carcinoma [65], prostate cancer[66], pancreatic cancer [67], gastric cancer [68] and colon cancer [69]. Overexpressed BGN has been reported to be associated with the aggressive growth and metastasis of tumors [68, 69], and with a worse prognosis for patients with gastric cancer [70] and pancreatic adenocarcinoma [71].These findings suggest that the BGN gene may act as either a potential therapeutic target or prognostic biomarker in multiple types of cancer. However, the transcriptional expression and prognostic value of the BGN gene in human cancers requires further investigation.

Module analysis from the PPI network showed that pancreatic ductal adenocarcinoma is closely related to Focal adhesion, PI3K-Akt signaling pathway and ECM-receptor interaction. Focal adhesion is a complex, dynamic process involving the driving activity of actin cytoskeleton and the participation of specific receptors and signal transduction [72]. Studies have found that focal adhesions are intensely involved in multiple key pathways of tumor migration and metastasis [73]. Research by Lu et al. showed that abnormal ECM can promote the growth and metastasis of tumors by directly promoting cell metastasis on the one hand, and indirectly by promoting the formation of tumor microvessels on the other hand [74]. It is noteworthy that 11 of the 17 genes in the module (COL1A1, COL3A1, COL1A2, LAMB3, ITGA3, COL11A1, LAMA3, COL5A2, FN1, COL6A3 and LAMC2) are involved in three pathways, which strengthens the findings of the COL family in pancreatic ductal adenocarcinoma.

To study the expression levels and prognostic value of 10 hub genes, we used GEPIA database and UALCAN for expression validation and prognostic analysis. The GEPIA database showed all the 10 hub genes are upregulated in PDAC compared to normal pancreas tissue. The results of the prognostic analysis showed that the upregulated expression of ITGA3 and MET significantly reduced the survival time of PDAC patients. Therefore, ITGA3 and MET appear to be ideal prognostic indicators for pancreatic ductal adenocarcinoma. In sum, we identified DEGs and performed GO analysis, pathway enrichment analysis, and PPI network construction to understand their roles in pancreatic ductal adenocarcinoma. In addition, we identified ITGA3 and MET as hub genes and evaluated their prognostic value. This study provided evidence for early diagnosis and prognostic evaluation of pancreatic ductal adenocarcinoma at the molecular level, but these findings need to be confirmed by subsequent laboratory studies.

**Conclusion:**

In this research, We employed bioinformatics tools to define the DEGs of Pancreatic ductal adenocarcinoma to inquire the biological and pathways, as well as screening and evaluating some hub genes. our studies offer some advice and references for the early detection and treatment of pancreatic ductal adenocarcinoma at the molecular level. we screened 10 hub genes from 202 DEGs under special conditions. These findings could help researchers better understand the genesis, patholoy, pathophysiology, and potential molecular pathways and gene targets of PDAC, which could lead to the development of diagnostic biomarkers and therapy options for the disease. However, our research requires laboratory studies, and we are very interested in having other experimental researchers follow this path to confirm the results presented in this study.

# Acknowledgements:

None

EMG and Clinical Neurophysiology; 1995 Oct 15-19; Kyoto, Japan. Amsterdam: Elsevier; 1996.

- **Conference paper:** Bengtsson S, Solheim BG. Enforcement of data protection, privacy and security in medical informatics. In: Lun KC, Degoulet P, Piemme TE, Rienhoff O, editors. MEDINFO 92. Proceedings of the 7th World Congress on Medical Informatics; 1992 Sep 6-10; Geneva, Switzerland.

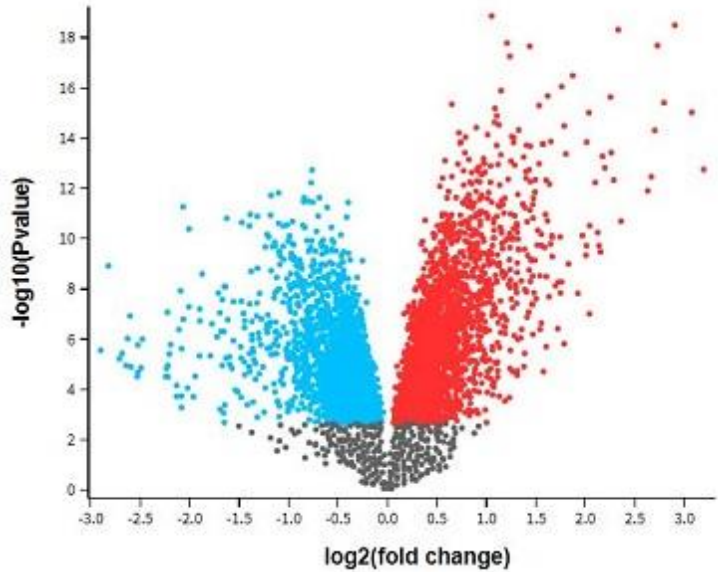

GSE28735

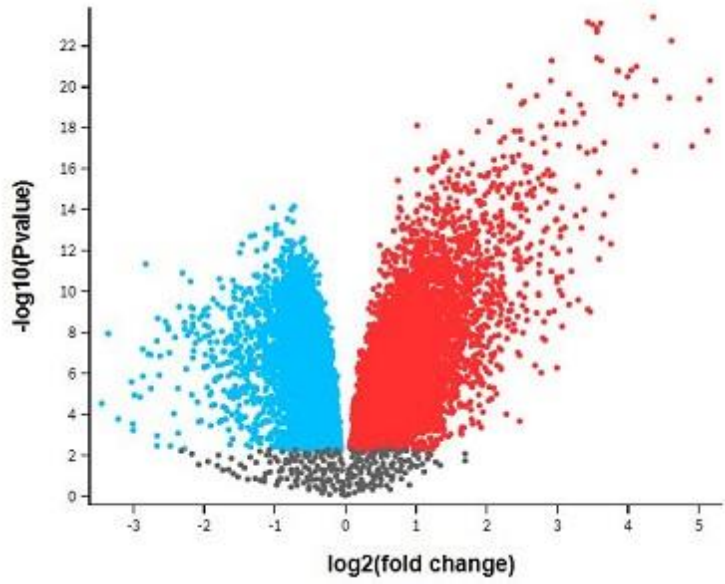

GSE15471

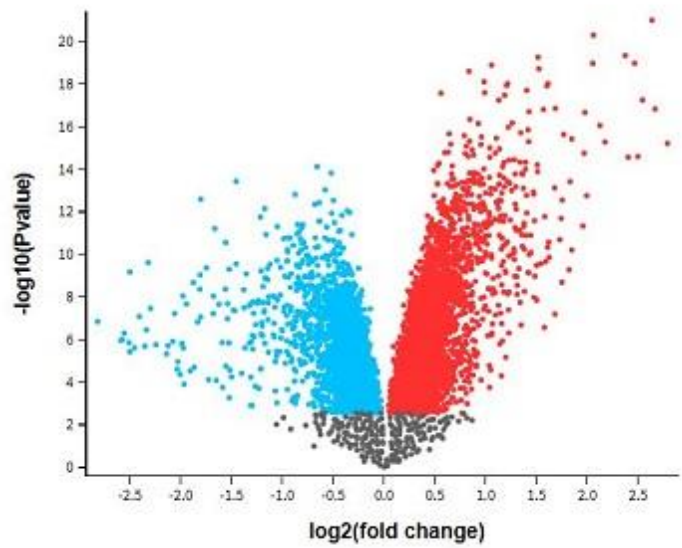

GSE62452

**Figure 1.** All significant DEGs show by Volcano diagrams which determine by considering adj. p-value and |log2FC| as the cutoff values for statistical analysis of each dataset, Red color means uDEGs, the blue color represents dDEGs, and Black color represents genes that are not significantly different in expression.

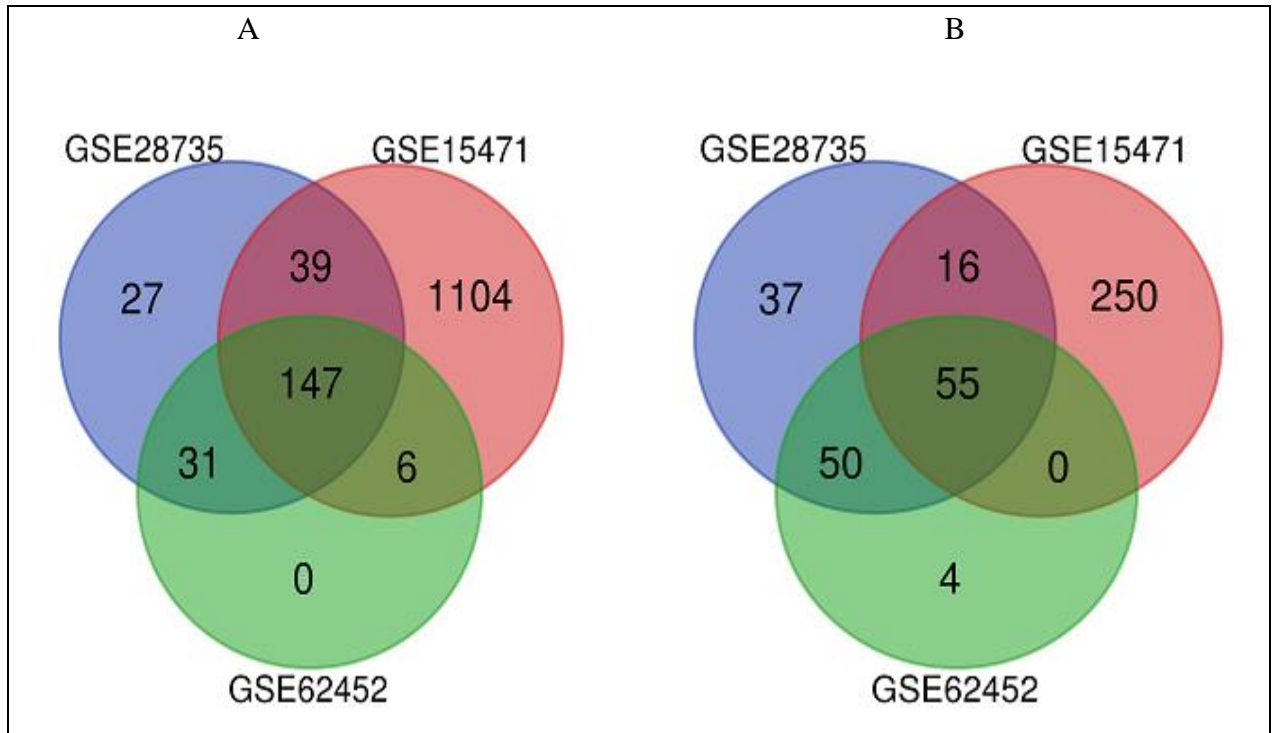

**Figure 2.** (A, B) The overlapping genes among DEGs of GSE28735, GSE15471 and GSE62452 microarrays show by Venn Diagrams. A total of 147 uDEGs and 55 dDEGs were found in the intersections.

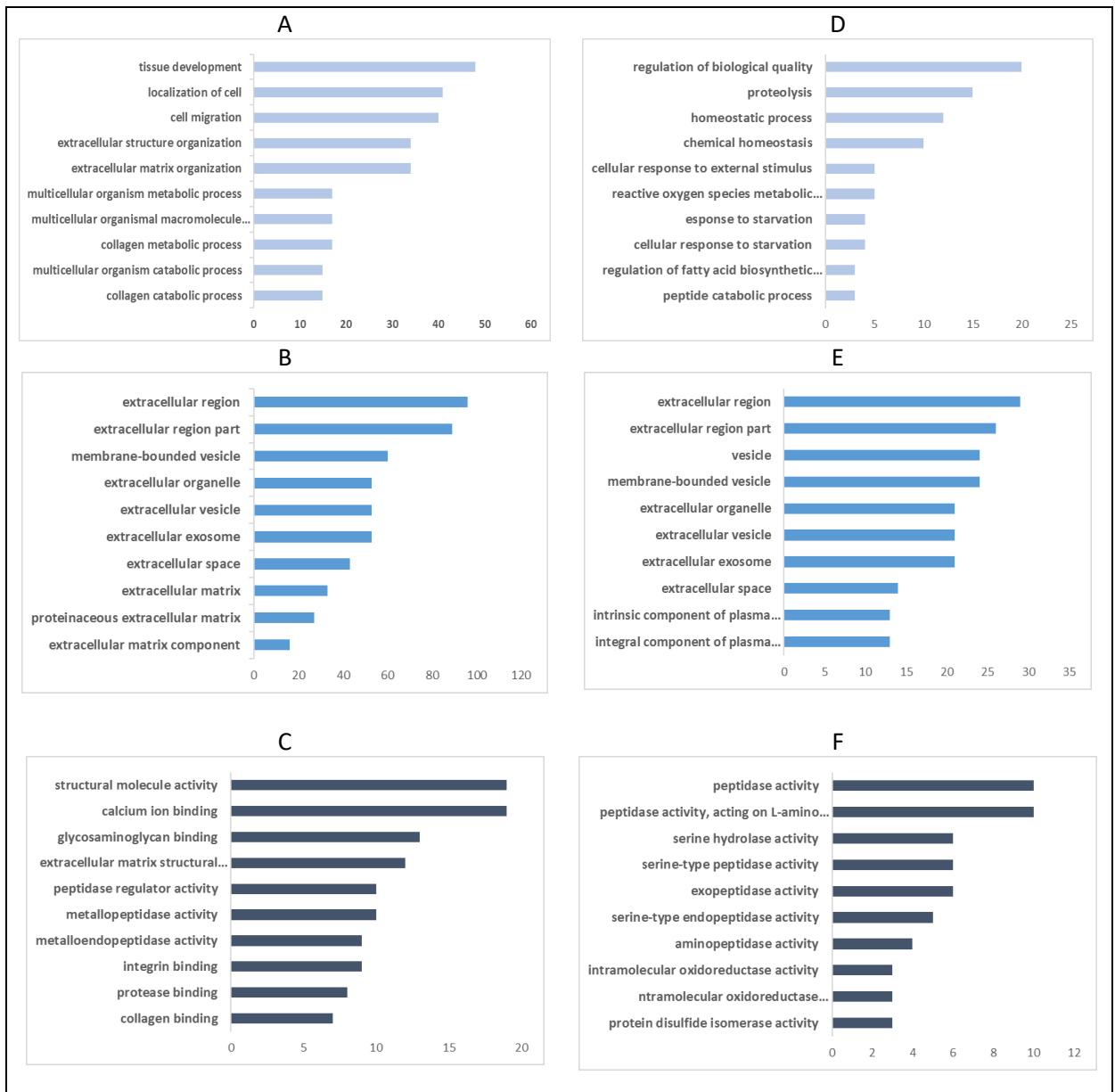

**Figure 3.** Gene ontology analysis of DEGs related to Pancreatic Ductal Adenocarcinoma. The x-axis stands for the number of DEGs, and the vertical axis stands for GO terms, (A) The top 10 enriched of 147 uDEGs : A ) biological processes (BP), B) cellular component (CC), C) molecular function (MF) .The top 10 enriched of 55 dDEGs : D) BP, E) CC, F) MF.

**Table 1.** DEGs in Pancreatic ductal adenocarcinoma shared in 3 microarrays.

| DEGS | Total | Elements |
|---|---|---|
| uDEGs | 147 | PLAU, STYK1, MMP7, IGF2BP3, SERPINB5, HEPH, CEACAM6, MET, LRRN1, ITGA3, TGM2, COL1A1, ANLN, DGKH, ADGRF1, FNDC1, SLC6A6, TRIM29, EDNRA, GALNT5, LTBP1, DUOX2, NRP2, MFAP5, EPHA4, CST1, PLA2R1, FN1, KRT17, IFI27, ARNTL2, LAMB3, TNFAIP6, LEF1, MYOF, ANO1, DPCR1, VSIG1, COL5A2, ANKRD22, DDX60, OSBPL3, TMPRSS4, ANTXR1, KYNU, CAPG, CCL20, SDR16C5, ADAM9, MATN3, OLR1, NPR3, SLPI, GPRC5A, NOX4, NT5E, CTSK, SULF2, IL1RAP, GPX2, ACSL5, CDH11, ITGA2, SCEL, FBN1, SLC6A14, BGN, LAMA3, KRT6A, MMP1, COL8A1, IGFBP5, MMP12, SLC2A1, CD109, SLC22A3, DHRS9, ADAMTS12, LAMP5, ECT2, MMP11, CDH3, COL3A1, TMEM45B, EDIL3, ASPM, FAP, INPP4B, INHBA, ANXA10, TCN1, POSTN, MMP14, PLAT, NQO1, ADAM28, ANXA8L1, SRPX2, IFI44L, CEACAM5, CEMIP, CTSE, OAS2, HK2, GABRP, TOP2A, MICAL2, COMP, AEBP1, SYTL2, THBS2, RUNX2, DKK1, KRT7, TSPAN1, KRT19, VCAN, SULF1, FERMT1, LAMC2, CXCL5, GCNT3, COL6A3, EGLN3, CST2, COL10A1, LCN2, ACTA2, PLAC8, CEACAM1, COL11A1, SERPINB3, ETV1, MLPH, FBXO32, TFF1, CLDN18, FGD6, COL1A2, CP, FXYD3, MBOAT2, SEMA3C, DPYSL3, COL12A1, AGR2, SLC44A4 |
| dDEGs | 55 | MT1G, BTG2, LMO3, BNIP3, ERO1B, CTNND2, GUCA1C, FGL1, RGN, PDIA2, C5, EGF, FAM3B, AQP8, PM20D1, CTRL, BACE1, SYCN, NRCAM, NUCB2, HOMER2, PNLIPRP1, KLK1, PAK3, LIFR, CHRM3, SLC17A4, TMED6, ALB, ERP27, KIAA1324, TRHDE, GNMT, SLC43A1, CELA2B, NR5A2, SERPINI2, AQP12B, GP2, GATM, SLC4A4, ACADL, IL22RA1, F8, DPP10, AOX1, GPHA2, PDK4, RBPJL, ANPEP, EPHX2, SLC39A5, NRG4, PAIP2B, F11 |

Table 2. Gene ontology analysis of uDEGs related to Pancreatic Ductal Adenocarcinoma.

| Category | Term | Count | % | P value | FDR |
|---|---|---|---|---|---|
| GOTERM_BP_FAT | GO:0030198~extracellular matrix organization | 34 | 23.12925 | 3.34E-26 | 5.57E-23 |
| GOTERM_BP_FAT | GO:0043062~extracellular structure organization | 34 | 23.12925 | 3.68E-26 | 5.57E-23 |
| GOTERM_BP_FAT | GO:0030574~collagen catabolic process | 15 | 10.20408 | 1.85E-16 | 1.87E-13 |
| GOTERM_BP_FAT | GO:0044243~multicellular organism catabolic process | 15 | 10.20408 | 8.59E-16 | 5.30E-13 |
| GOTERM_BP_FAT | GO:0032963~collagen metabolic process | 17 | 11.56463 | 8.77E-16 | 5.30E-13 |
| GOTERM_BP_FAT | GO:0044259~multicellular organismal macromolecule metabolic process | 17 | 11.56463 | 1.92E-15 | 9.70E-13 |
| GOTERM_BP_FAT | GO:0044236~multicellular organism metabolic process | 17 | 11.56463 | 1.99E-14 | 8.61E-12 |
| GOTERM_BP_FAT | GO:0016477~cell migration | 40 | 27.21088 | 5.70E-14 | 2.16E-11 |
| GOTERM_BP_FAT | GO:0009888~tissue development | 48 | 32.65306 | 1.59E-13 | 5.35E-11 |
| GOTERM_BP_FAT | GO:0051674~localization of cell | 41 | 27.89116 | 4.78E-13 | 1.31E-10 |
| GOTERM_CC_FAT | GO:0044421~extracellular region part | 89 | 60.54422 | 2.71E-25 | 8.35E-23 |
| GOTERM_CC_FAT | GO:0005576~extracellular region | 96 | 65.30612 | 8.17E-25 | 1.26E-22 |
| GOTERM_CC_FAT | GO:0031012~extracellular matrix | 33 | 22.44898 | 1.28E-19 | 1.32E-17 |
| GOTERM_CC_FAT | GO:0005578~proteinaceous extracellular matrix | 27 | 18.36735 | 6.89E-18 | 5.31E-16 |
| GOTERM_CC_FAT | GO:0005615~extracellular space | 43 | 29.2517 | 4.55E-14 | 2.80E-12 |
| GOTERM_CC_FAT | GO:0044420~extracellular matrix component | 16 | 10.88435 | 8.59E-14 | 4.41E-12 |
| GOTERM_CC_FAT | GO:0070062~extracellular exosome | 53 | 36.05442 | 8.29E-10 | 3.44E-08 |

| Category | Term | Count | % | P-Value | Benjamini |
|---|---|---|---|---|---|
| GOTERM_CC_FAT | GO:1903561~extracellular vesicle | 53 | 36.05442 | 9.92E-10 | 3.44E-08 |
| GOTERM_CC_FAT | GO:0043230~extracellular organelle | 53 | 36.05442 | 1.00E-09 | 3.44E-08 |
| GOTERM_CC_FAT | GO:0031988~membrane-bounded vesicle | 60 | 40.81633 | 4.33E-09 | 1.33E-07 |
| GOTERM_MF_FAT | GO:0005201~extracellular matrix structural constituent | 12 | 8.163265 | 6.22E-11 | 3.41E-08 |
| GOTERM_MF_FAT | GO:0005539~glycosaminoglycan binding | 13 | 8.843537 | 1.39E-07 | 3.82E-05 |
| GOTERM_MF_FAT | GO:0005178~integrin binding | 9 | 6.122449 | 2.50E-06 | 4.57E-04 |
| GOTERM_MF_FAT | GO:0004222~metalloendopeptidase activity | 9 | 6.122449 | 4.05E-06 | 5.54E-04 |
| GOTERM_MF_FAT | GO:0005518~collagen binding | 7 | 4.761905 | 1.36E-05 | 0.00149 |
| GOTERM_MF_FAT | GO:0005509~calcium ion binding | 19 | 12.92517 | 2.22E-05 | 0.00203 |
| GOTERM_MF_FAT | GO:0008237~metallopeptidase activity | 10 | 6.802721 | 2.64E-05 | 0.002065 |
| GOTERM_MF_FAT | GO:0002020~protease binding | 8 | 5.442177 | 4.92E-05 | 0.003368 |
| GOTERM_MF_FAT | GO:0005198~structural molecule activity | 19 | 12.92517 | 6.88E-05 | 0.004191 |
| GOTERM_MF_FAT | GO:0061134~peptidase regulator activity | 10 | 6.802721 | 8.25E-05 | 0.004334 |

**Table 2 continued.** Gene ontology analysis of dDEGs related to Pancreatic Ductal Adenocarcinoma.

| Category | Term | Count | % | P value | FDR |
|---|---|---|---|---|---|
| GOTERM_BP_FAT | GO:0006508~proteolysis | 15 | 27.27273 | 3.79E-04 | 0.635703 |
| GOTERM_BP_FAT | GO:0048878~chemical homeostasis | 10 | 18.18182 | 0.003956 | 1 |
| GOTERM_BP_FAT | GO:0043171~peptide catabolic process | 3 | 5.454545 | 0.003975 | 1 |
| GOTERM_BP_FAT | GO:0072593~reactive oxygen species metabolic process | 5 | 9.090909 | 0.00457 | 1 |
| GOTERM_BP_FAT | GO:0042304~regulation of fatty acid biosynthetic process | 3 | 5.454545 | 0.004768 | 1 |
| GOTERM_BP_FAT | GO:0009267~cellular response to starvation | 4 | 7.272727 | 0.005131 | 1 |
| GOTERM_BP_FAT | GO:0065008~regulation of biological quality | 20 | 36.36364 | 0.006057 | 1 |
| GOTERM_BP_FAT | GO:0071496~cellular response to external stimulus | 5 | 9.090909 | 0.007625 | 1 |
| GOTERM_BP_FAT | GO:0042592~homeostatic process | 12 | 21.81818 | 0.009278 | 1 |
| GOTERM_BP_FAT | GO:0042594~response to starvation | 4 | 7.272727 | 0.011222 | 1 |
| GOTERM_CC_FAT | GO:0005576~extracellular region | 29 | 52.72727 | 2.69E-05 | 0.003033 |
| GOTERM_CC_FAT | GO:0044421~extracellular region part | 26 | 47.27273 | 3.55E-05 | 0.003033 |
| GOTERM_CC_FAT | GO:0070062~extracellular exosome | 21 | 38.18182 | 8.56E-05 | 0.003149 |

| Category | Term | Count | % | PValue | Benjamini |
|---|---|---|---|---|---|
| GOTERM_CC_FAT | GO:1903561~extracellular vesicle | 21 | 38.18182 | 9.19E-05 | 0.003149 |
| GOTERM_CC_FAT | GO:0043230~extracellular organelle | 21 | 38.18182 | 9.24E-05 | 0.003149 |
| GOTERM_CC_FAT | GO:0031988~membrane-bounded vesicle | 24 | 43.63636 | 1.10E-04 | 0.003149 |
| GOTERM_CC_FAT | GO:0031982~vesicle | 24 | 43.63636 | 2.04E-04 | 0.00498 |
| GOTERM_CC_FAT | GO:0005615~extracellular space | 14 | 25.45455 | 2.35E-04 | 0.005026 |
| GOTERM_CC_FAT | GO:0005887~integral component of plasma membrane | 13 | 23.63636 | 0.002554 | 0.048524 |
| GOTERM_CC_FAT | GO:0031226~intrinsic component of plasma membrane | 13 | 23.63636 | 0.003544 | 0.054511 |
| GOTERM_MF_FAT | GO:0008238~exopeptidase activity | 6 | 10.90909 | 1.88E-05 | 0.005976 |
| GOTERM_MF_FAT | GO:0070011~peptidase activity, acting on L-amino acid peptides | 10 | 18.18182 | 2.04E-04 | 0.028008 |
| GOTERM_MF_FAT | GO:0008233~peptidase activity | 10 | 18.18182 | 2.64E-04 | 0.028008 |
| GOTERM_MF_FAT | GO:0004177~aminopeptidase activity | 4 | 7.272727 | 3.62E-04 | 0.028754 |
| GOTERM_MF_FAT | GO:0008236~serine-type peptidase activity | 6 | 10.90909 | 0.001439 | 0.076014 |
| GOTERM_MF_FAT | GO:0017171~serine hydrolase activity | 6 | 10.90909 | 0.001508 | 0.076014 |
| GOTERM_MF_FAT | GO:0003756~protein disulfide isomerase activity | 3 | 5.454545 | 0.001912 | 0.076014 |
| GOTERM_MF_FAT | GO:0016864~intramolecular oxidoreductase activity, transposing S-S bonds | 3 | 5.454545 | 0.001912 | 0.076014 |
| GOTERM_MF_FAT | GO:0004252~serine-type endopeptidase activity | 5 | 9.090909 | 0.006795 | 0.240095 |
| GOTERM_MF_FAT | GO:0016860~intramolecular oxidoreductase activity | 3 | 5.454545 | 0.011153 | 0.354667 |

**Table 3.** KEGG pathway analysis of DEGs associated with pancreatic ductal adenocarcinoma.

| Expression | Term | Count | % | P value | FDR | Genes |
|---|---|---|---|---|---|---|
| Upregulated | hsa04510: Focal adhesion | 15 | 10.20408 | 1.09E-08 | 6.24E-07 | LAMB3, ITGA3, COL11A1, ITGA2, LAMA3, FN1, LAMC2, THBS2, COL1A1, COMP, COL3A1, COL1A2, COL5A2, COL6A3, MET |
| | hsa04151: PI3K-Akt signaling pathway | 15 | 10.20408 | 6.11E-06 | 1.76E-04 | LAMB3, ITGA3, COL11A1, ITGA2, LAMA3, FN1, LAMC2, THBS2, COL1A1, COMP, COL3A1, COL1A2, COL5A2, COL6A3, MET |
| | hsa04512: ECM-receptor interaction | 14 | 9.52381 | 1.83E-12 | 2.11E-10 | LAMB3, ITGA3, COL11A1, ITGA2, LAMA3, FN1, LAMC2, THBS2, COL1A1, COMP, COL3A1, COL1A2, COL5A2, COL6A3 |
| Downregulated | hsa04972: Pancreatic secretion | 5 | 9.090909 | 7.45E-04 | 0.059593 | PNLIPRP1, CHRM3, CELA2B, CTRL, SLC4A4 |
| | hsa04610: Complement and coagulation cascades | 3 | 5.454545 | 0.038246 | 1 | C5, F8, F11 |

**Table 3 continued.** KEGG pathway analysis of DEGs associated with pancreatic ductal adenocarcinoma.

| Expression | Term | Count | % | P value | FDR | Genes |
|---|---|---|---|---|---|---|
| | hsa04012: ErbB signaling pathway | 3 | 5.454545 | 0.058032 | 1 | NRG4, EGF, PAK3 |

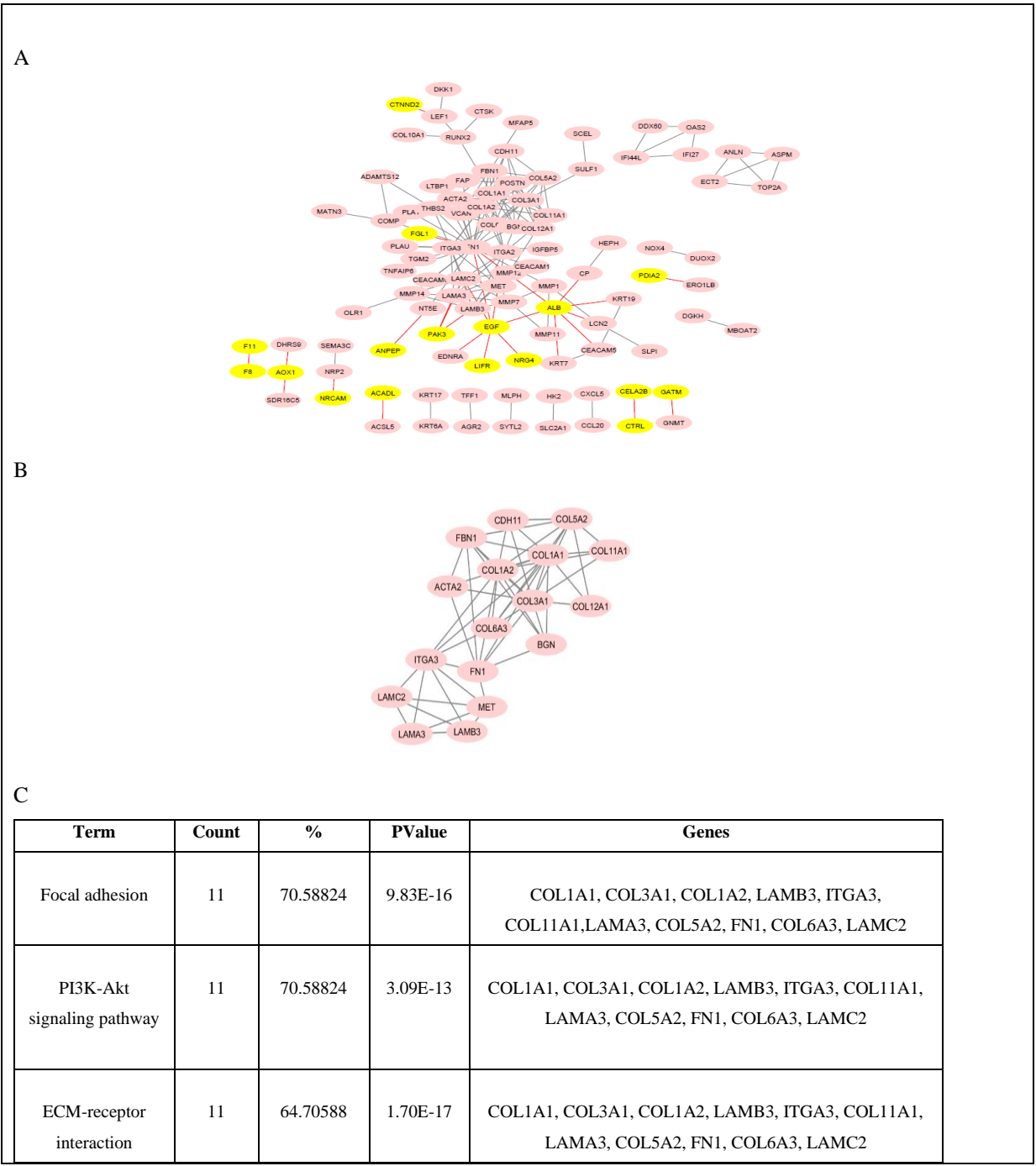

**Figure 4**. PPI network construction, module analysis, and pathway enrichment analysis. Protein-protein interaction network for products of DEGs. A total of 202 nodes and 174 interaction associations were identified. (A) The nodes mean proteins; the edges mean the interactions of proteins; pink circles meant uEGs and yellow circles meant dEGs (B) Module analysis

based on Cytoscape software. (C) KEGG pathway enrichment analysis of DEGs in the module.

**Table 4.** Connectivity and regulation of the top 10 hub genes.

| Gene symbol | Gene title | Connectivity | Regulation |
|---|---|---|---|
| COL1A1 | collagen type I alpha 1 chain | 24 | Up |
| COL3A1 | collagen type III alpha 1 chain | 22 | Up |
| COL1A2 | collagen type I alpha 2 chain | 20 | Up |
| FN1 | fibronectin 1 | 16 | Up |
| COL5A2 | collagen type V alpha 2 chain | 16 | Up |
| ITGA3 | integrin subunit alpha 3 | 16 | Up |
| FBN1 | fibrillin 1 | 14 | Up |
| MET | MET proto-oncogene, receptor tyrosine kinase | 10 | Up |
| COL6A3 | collagen type VI alpha 3 chain | 10 | Up |
| BGN | biglycan | 10 | Up |

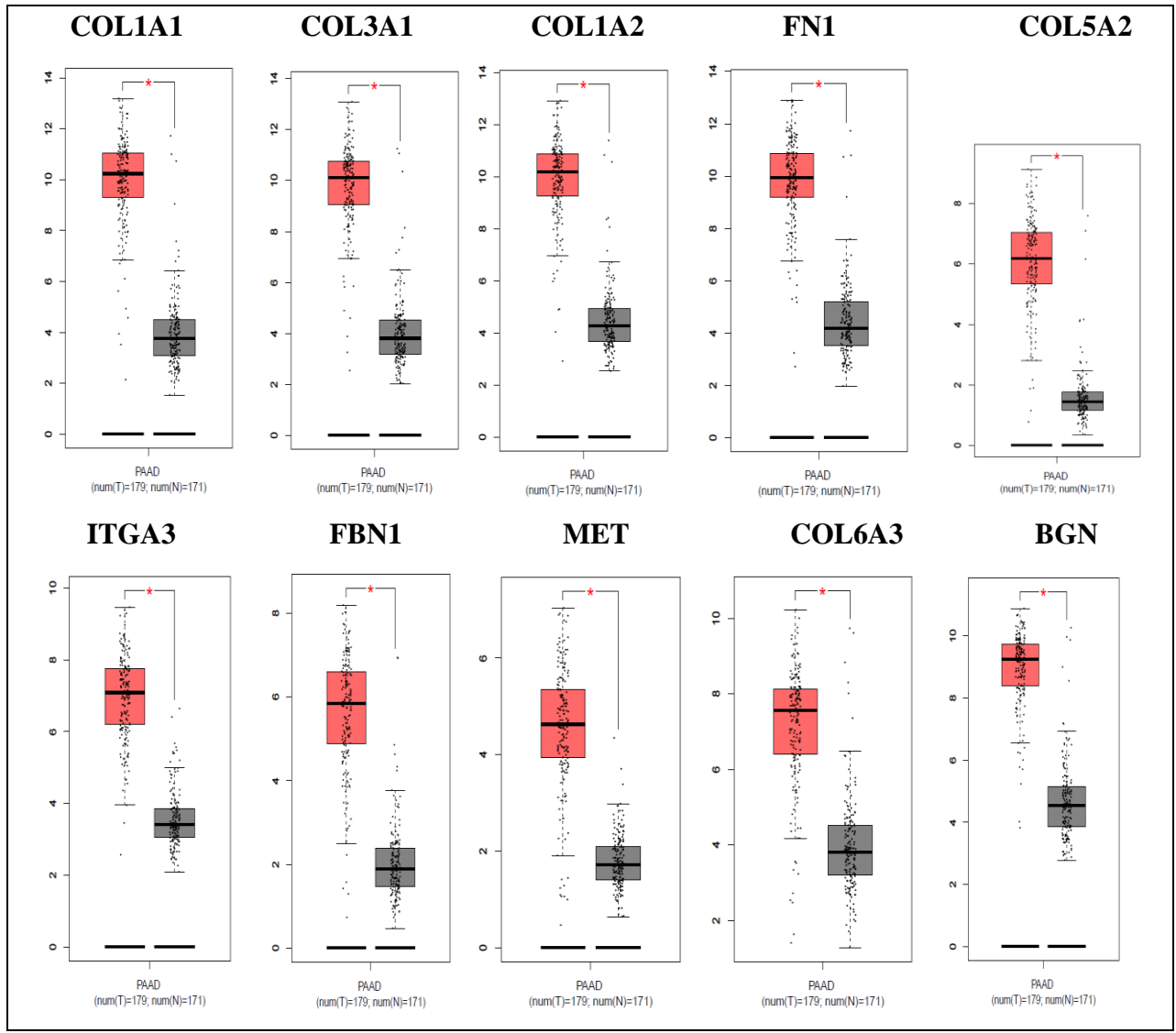

**Figure 5.** Validation of the gene expression levels of COL1A1, COL3A1, COL1A2, FN1, COL5A2, ITGA3, FBN1, MET, COL6A3, and BGN between PDAC and normal pancreatic tissues in the GEPIA database. They are significantly upregulated in PDAC compared with normal tissues (P<0.01). The red * represents P<0.01.

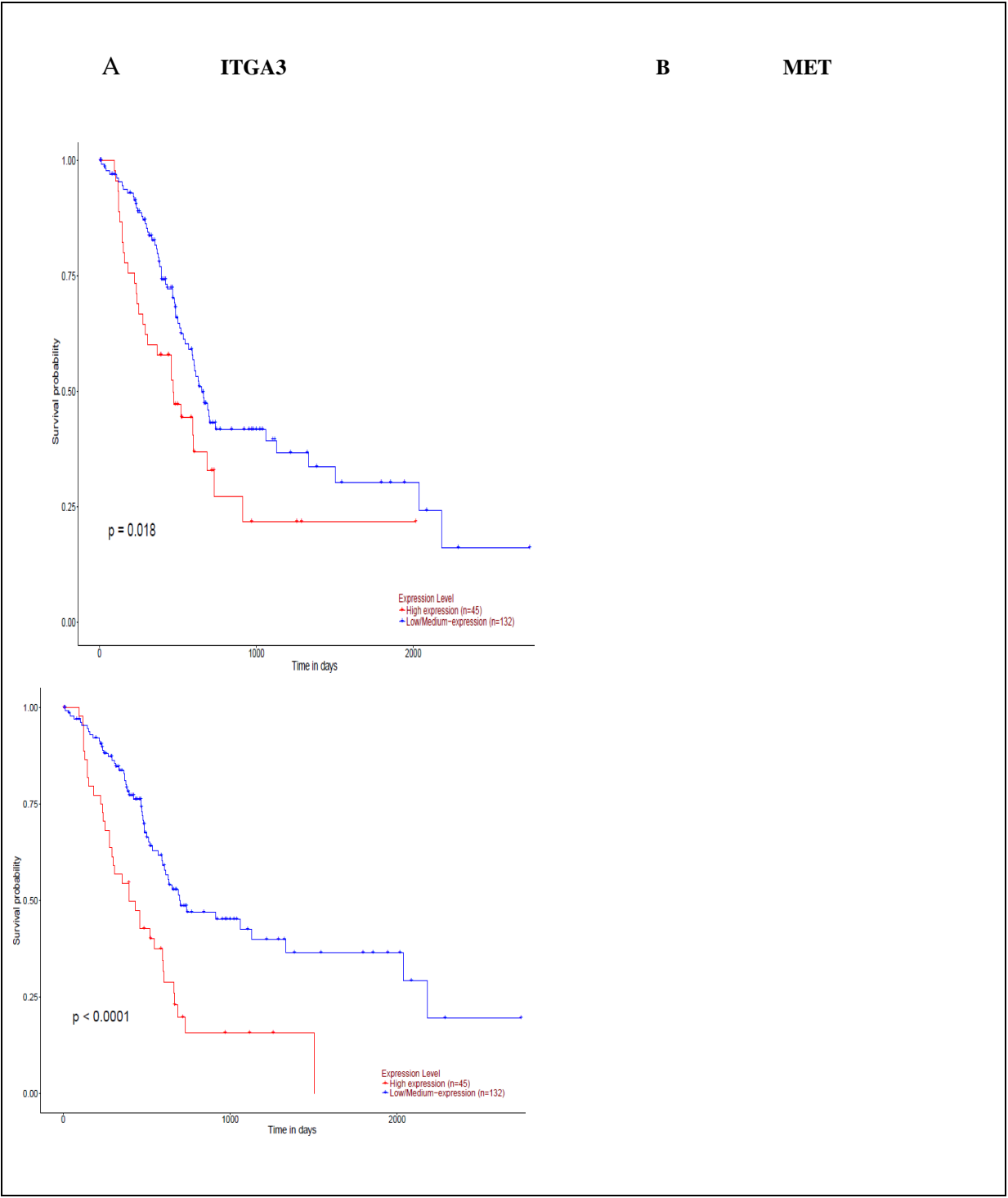

**Figure 6.** UALCAN overall survival analysis plot of the top 10 hub genes expressed in pancreatic ductal adenocarcinoma samples patient samples and 2 DEGs among the top 10 hub genes

that are significantly related to the survival of pancreatic ductal adenocarcinoma patients (P<0.05). (A) ITGA3; (B) MET